# Nonrelativistic energy of tritium-containing hydrogen molecule isotopologues

Krzysztof Pachucki[a] and Jacek Komasa[b]

[a]Faculty of Physics, University of Warsaw, Pasteura 5, 02-093 Warsaw, Poland;
[b]Faculty of Chemistry, Adam Mickiewicz University, Uniwersytetu Poznańskiego 8, 61-614 Poznań, Poland

**ABSTRACT**
The nonrelativistic energy of low lying rovibrational levels of HT, DT, and $T_2$ is determined to an absolute accuracy of $10^{-7} - 10^{-8}$ cm$^{-1}$ using the variational method with the four-body nonadiabatic James-Coolidge functions. The new results increase the accuracy of the nonrelativistic component of the energy levels by several orders of magnitude. As a consequence, the total transition energies are improved by at least an order of magnitude.

**KEYWORDS**
Hydrogen molecule, tritium isotopologues, nonrelativistic energy, nonadiabatic James-Coolidge wavefunction;

## 1. Introduction

The hydrogen molecule exists as six isotopic species. Their electronic structures differ mainly because of variations in nuclear masses but also, to a much lesser extent, due to differences in nuclear magnetic moments and charge radii. Three of the isotopologues, containing stable nuclei, were thoroughly studied both experimentally and theoretically, contributing valuable information on the tiniest details of their electronic structure [1–22]. The remaining radioactive isotopologues where studied experimentally much less frequently for obvious technical and safety reasons. Nevertheless, their first observations were performed in 1949 by Dieke and Tomkins [23], whenever sufficient quantities of the radioactive material became accessible. After a decade of studies, exhaustive spectra of all tritium-containing isotopic species were determined [24] with an accuracy of $0.01 - 0.02$ cm$^{-1}$, becoming a reference for a long time. In the next half-century, the only reports of observations of spectra in tritiated molecular hydrogen are from Edwards *et al.* [25, 26], Viers and Rosenblatt [27], and Chuang and Zare [28], the latter providing data with an estimated accuracy greater than $0.01$ cm$^{-1}$.

A strong impetus to study both experimentally and theoretically the T-isotopologues of $H_2$ comes from the Karlsruhe Tritium Neutrino (KATRIN) project, running since 2001. Its primary purpose is to measure the absolute neutrino mass by studying the end point of the $\beta$-decay kinematics of the molecular tritium $T_2$ [29]

---

within the radioactive process $T_2 \to {}^3\text{HeT}^+ + e + \bar{\nu}_e$. Initially, however, the $\beta$-decay of the triton placed in a molecular surrounding was studied without reference to the neutrino mass issue. In the 1950s Schwartz [30] and Cantwell [31] made the first theoretical estimation of the probability of molecular transitions accompanying the radioactive process, but the first accurate calculations on HeH$^+$ in the context of triton decay in HT were performed by Wolniewicz in 1965 [32]. Interest in this process increased rapidly when the non-zero mass of the neutrino was postulated, and the $\beta$-decay process in tritium has been proposed as the best method of its determination. In the 1980s, Kołos *et al.* addressed this issue from the theoretical side in a series of papers [33–40].

There has been increased interest in the spectroscopy of tritium isotopologues of the hydrogen molecule in recent years. Researchers of Vrije Universiteit Amsterdam and Tritium Laboratory Karlsruhe overcame the technical difficulties related to handling the radioactive tritium samples and presented new, more accurate measurements of the transition energies in HT, DT, and $T_2$. Schloesser *et al.* [41] reported on the determination of $T_2$ Q(0 − 5) transition energies in the fundamental vibrational band of the ground electronic state. Using high-resolution coherent anti-Stokes Raman spectroscopy (CARS), the uncertainty of 0.02 cm$^{-1}$ was reached. Shortly after that, results with a 50-fold increase in precision was obtained by Trivikram *et al.* [42]. For DT, Lai *et al.* [43] performed measurements of Q(0 − 5) transitions, achieving an accuracy of $< 0.0004$ cm$^{-1}$, in agreement with even more accurate theoretical predictions. Most recently, analogous measurements have been performed for HT at similar accuracy [44].

In contrast to measurements, theoretical studies of tritium isotopologues of H$_2$ are not burdened with difficulties related to the radioactivity of the sample and, in principle, the difficulty level of such computational studies is not much higher than in the case of H$_2$. Therefore, despite the lack of accurate experimental references, the theoretical data of increasing accuracy are well known in the literature [45–54]. The first accurate computations on HT, DT, and $T_2$ can be attributed to Wolniewicz, who in the 1990s accounted for nonadiabatic and relativistic corrections [45, 46] to the Born-Oppenheimer energy. Currently, the most accurate rovibrational energy levels can be obtained by means of the freely available computer program H2Spectre [55]. Their absolute accuracy is of the order of $10^{-3} - 10^{-4}$ cm$^{-1}$ and is limited, in the first place, by neglecting $\mathcal{O}(\mu_n^{-3})$ terms ($\mu_n$ is the reduced mass of nuclei) in the nonrelativistic energy and, in the second place, by the neglected quantum electrodynamic (QED) recoil corrections.

From the theoretical point of view, the recent announcements [44] of a further significant increase in the accuracy of the measurements are important because they motivate further reduction of the uncertainty of our calculations on tritiated species. Another important motivation is the search for the New Physics [43, 57], or more precisely, the search for yet unknown interactions at the atomic scale, which relies on the comparison of precise theoretical results with experimental data.

In general, the most accurate contemporary measurements and calculations are in good agreement [1, 3–5, 11–13, 17, 41, 43, 58–74]. Nonetheless, there are exceptional cases when experimental data differ slightly with each other or when theoretical results disagree with observations. One instance of such a disagreement concerns the $R(1)$ line of the first overtone in the HD molecule. Among four independent experimental results, three are consistent with each other [6, 14, 22] and one deviates from them [20]. In addition, the theoretical prediction [10] disagrees by $2\sigma$ with all of them. Another interesting example of such a disagreement has been noted for the dissociation energy of HD. The experimental value of $D_0 = 36\,405.783\,66(36)\,\text{cm}^{-1}$ [75] differs by $3\sigma$ from the calculated $D_0 = 36\,405.782\,477(26)\,\text{cm}^{-1}$ [16]. In view of the perfect agreement existing for $\text{H}_2$ and $\text{D}_2$, the question of whether these discrepancies are specific to heteronuclear species becomes justified. Reported in this work new data for the other heteronuclear isotopologues will contribute to addressing this question.

An accurate solution of a molecular Schrödinger equation with Coulomb interactions is one of the fundamental problems of quantum chemistry. The complexity of the Schrödinger equation prevents its exact solution in a general multiparticle case and enforces approximations to be made. The most common approximation applied in solving the molecular time-independent Schrödinger equation is the separation of the electronic and nuclear variables, often referred to as the adiabatic or Born-Oppenheimer approximation. The error resulting from this separation can be partially compensated later by adding adiabatic and nonadiabatic corrections. The main objective of this work is an application of the variational method, based on four-body exponential wavefunctions, without introducing the separation of the variables, to tritium-containing isotopologues of the hydrogen molecule. Such an approach accounts for complete finite mass (nonadiabatic) effects. The first calculations have already been performed for low lying energy levels of $\text{H}_2$, HD, and $\text{D}_2$ [76–79]. This method enabled determination of the eigenvalues of the four-body Schrödinger equation with the relative accuracy of $10^{-12}$, which, e.g., for $\text{H}_2$ corresponds to $10^{-7}\,\text{cm}^{-1}$ in terms of the nonrelativistic dissociation energy. This accuracy is preserved for rotationally excited levels and is by 2-3 orders of magnitude higher than any previous result available for the hydrogen molecule. Obtaining such a level of accuracy is possible only if the coupling between the rotational and electronic angular momentum is included. Therefore, a crucial part of our method relies on a construction and implementation of the wavefunction which effectively accounts for this coupling. Apart from the accurate energy spectrum, a collection of very accurate nonadiabatic wavefunctions is obtained. These functions will be used in future applications to determine various molecular properties.

## 2. The nonadiabatic wavefunction

The theoretical method described here is relevant to molecules consisting of two electrons 1 and 2, and two nuclei $A$ and $B$. The nuclei feature masses $m_A$ and $m_B$ and charges $Z_A$ and $Z_B$. The nonrelativistic Hamiltonian for such a system, expressed in atomic units (a.u.) and with commonly used symbols, reads

$$\hat{H} = -\frac{1}{2\,m_A}\nabla_A^2 - \frac{1}{2\,m_B}\nabla_B^2 - \frac{1}{2}\nabla_1^2 - \frac{1}{2}\nabla_2^2 + \frac{1}{r_{12}} + \frac{Z_A Z_B}{r_{AB}} - \frac{Z_A}{r_{1A}} - \frac{Z_A}{r_{2A}} - \frac{Z_B}{r_{1B}} - \frac{Z_B}{r_{2B}}. \tag{1}$$

The Schrödinger equation $\hat{H}\Psi = E\,\Psi$ is solved directly, i.e., without separation of the electronic and nuclear degrees of freedom, using the variational method. The solution accounts for all the nonadiabatic effects and yields directly the nonrelativistic energy of a rovibrational level. We assign this approach with the acronym DNA, which stands for **d**irect **n**on**a**diabatic.

The nonadiabatic wavefunction $\Psi^{J,M}$ of a rotational level $J$ depends formally also on the quantum number $M$, which is the projection of $\vec{J}$ on the axis $Z$ of the laboratory frame. Because the rotational angular momentum of nuclei couples to the electronic angular momentum $\vec{L}$, forming the total angular momentum $\vec{J}$ of the molecule, the wavefunction should account for this coupling by involving components relevant to appropriate electronic states. The quantum number $\Lambda$ – the eigenvalue of the $\vec{n} \cdot \vec{L}$ operator and the inversion symmetry symbol $g$ or $u$ (for gerade or ungerade) are employed to distinguish between such states: $\Sigma_{g,u}$, $\Pi_{g,u}$, $\Delta_{g,u}$, ... . The most general wavefunction in the DNA approach is represented as a sum of components with growing $\Lambda$

$$\Psi^{J,M} = \Psi^{J,M}_{\Sigma_g} + \Psi^{J,M}_{\Sigma_u} + \Psi^{J,M}_{\Pi_g} + \Psi^{J,M}_{\Pi_u} + \Psi^{J,M}_{\Delta_g} + \Psi^{J,M}_{\Delta_u} + \ldots . \tag{2}$$

For a homonuclear molecule in the electronic $\Sigma_g^+$ state, this expansion is simplified by rejecting all the ungerade functions $\Psi^{J,M}_{\Lambda_u}$. Subsequent terms of the expansion (2) have the following explicit form

$$\Psi^{J,M}_{\Sigma_{g,u}} = \mathcal{Y}^J_M \, \Phi^J_{\Sigma_{g,u}} \qquad \text{for } J \geq 0 \tag{3}$$

$$\Psi^{J,M}_{\Pi_{g,u}} = \sqrt{\frac{2}{J(J+1)}} R \, \rho^i \left( \nabla^i_R \mathcal{Y}^J_M \right) \Phi^J_{\Pi_{g,u}} \qquad \text{for } J \geq 1 \tag{4}$$

$$\Psi^{J,M}_{\Delta_{g,u}} = \sqrt{\frac{4}{(J-1)J(J+1)(J+2)}} R^2 (\rho^i \rho'^j)^{(2)} \left( \nabla^i_R \nabla^j_R \mathcal{Y}^J_M \right) \Phi^J_{\Delta_{g,u}} \qquad \text{for } J \geq 2 \tag{5}$$

and so on for the higher electronic angular momenta. In the above equations we use the following notation

$$(\rho^i \rho'^j)^{(2)} \equiv \frac{1}{2} \left( \rho^i \rho'^j + \rho^j \rho'^i - \delta^{ij}_\perp \vec{\rho} \cdot \vec{\rho}' \right), \tag{6}$$

where $\vec{\rho}, \vec{\rho}' \equiv \vec{\rho}_1$ or $\vec{\rho}_2$, $\rho^i_a = \delta^{ij}_\perp r^j_{aA}$, $\delta^{ij}_\perp \equiv \delta^{ij} - n^i n^j$, and $n^i \equiv R^i/R$, with the Einstein summation convention assumed. The symbol $\mathcal{Y}^J_M = R^J Y^J_M(\vec{n})$ denotes a solid harmonic, and the normalization coefficients are such that the matrix elements of a scalar operator $Q$ involving electronic variables fulfill

$$\langle \Psi^{J,M}_\Sigma | Q | \Psi^{J,M}_\Sigma \rangle = \langle R^J \Phi^J_\Sigma | Q | R^J \Phi^J_\Sigma \rangle, \tag{7}$$

$$\langle \Psi^{J,M}_\Pi | Q | \Psi^{J,M}_\Pi \rangle = \langle R^J \Phi^J_\Pi | \vec{\rho} \, Q \, \vec{\rho} | R^J \Phi^J_\Pi \rangle, \tag{8}$$

$$\langle \Psi^{J,M}_\Delta | Q | \Psi^{J,M}_\Delta \rangle = \langle R^J \Phi^J_\Delta | (\rho^i \rho'^j)^{(2)} \, Q \, (\rho^i \rho'^j)^{(2)} | R^J \Phi^J_\Delta \rangle. \tag{9}$$

The functions $\Phi^J_\Lambda$ represent linear expansions

$$\Phi^J_\Lambda = \sum_{\{k\}} c_{\{k\}} \, (1 + \mathcal{P}_{12}) \, \Phi_{\{k\}} \tag{10}$$

in the following four-particle nonadiabatic James-Coolidge (naJC) basis functions

$$\Phi_{\{k\}} = e^{-\alpha R - \beta(\zeta_1 + \zeta_2)} \, R^{k_0} \, r^{k_1}_{12} \, \eta^{k_2}_1 \, \eta^{k_3}_2 \, \zeta^{k_4}_1 \, \zeta^{k_5}_2 \tag{11}$$

with $\zeta_1 = r_{1A} + r_{1B}$, $\eta_1 = r_{1A} - r_{1B}$, $\zeta_2 = r_{2A} + r_{2B}$, $\eta_2 = r_{2A} - r_{2B}$, and $\vec{R} = \vec{r}_{AB}$. $\alpha$ and $\beta$ denote nonlinear variational parameters, and $k_i$ are non-negative integers collectively denoted as $\{k\}$. The naJC basis functions having the same $\alpha$ and $\beta$ parameters form the so-called 'sector'. If needed, two or more sectors (with different pairs $\alpha^{(i)}$ and $\beta^{(i)}$) can be used. The symbol $\mathcal{P}_{12}$ in Eq. (10) denotes the electron permutation operator. The basis functions with $k_2 + k_3$ even (odd) have the subscript $g$ ($u$). For each pair $J$ and $L$, the function $\Phi_L^J$ has its own set of nonlinear parameters, so that $\Phi_{\{k\}}$ of Eq. (11) implicitly depend on indices $J$ and $L$. Finally, the linear coefficients $c_{\{k\}}$ are determined variationally by solving an eigenvalue problem by the inverse iteration method in extended-precision arithmetic.

The trial four-particle wavefunction defined above depends on interparticle distances only, i.e., it is translationally invariant and our results do not depend on the choice of the origin of the reference frame. We found it convenient for evaluation of matrix elements to locate the origin of the space-fixed coordinate system at the internuclear midpoint $\vec{O} = (\vec{R}_A + \vec{R}_B)/2$. Details on the wavefunction properties, on evaluation of the matrix elements, and on solving the general symmetric eigenvalue problem were described in [76–79] and will not be repeated here.

## 3. Results and discussion

In this section we present results obtained for HT, DT, and $T_2$ molecules using the DNA method applied to the naJC basis. The range of the considered rovibrational levels is limited to the three lowest vibrational states ($v = 0, 1, 2$) and six lowest rotational levels ($J = 0, \ldots, 5$). These results are to be used to predict the energy gaps between selected levels and compared with available spectroscopic data [41, 42, 44]. The choice of the levels was dictated by the available experimental data.

The physical constants employed in our calculations come from the 2018 CODATA compilation [81] and are assembled in Table 1.

**Table 1.** The physical constants [81] used in the present calculations: mass of proton, $m_p$, deuteron, $m_d$, and triton, $m_t$, as well as the Rydberg constant $\mathcal{R}_\infty$.

| Constant | Value |
|---|---|
| $m_p$ | 1836.152 673 43(11) a.u. |
| $m_d$ | 3670.482 967 88(13) a.u. |
| $m_t$ | 5496.921 535 73(27) a.u. |
| $2\mathcal{R}_\infty$ | 219 474.631 363 20(42) cm$^{-1}$ |

### 3.1. Uncertainty estimation

In order to determine the final extrapolated value of the energy as well as its uncertainty for all the individual rovibrational levels reported here, we performed a detailed study of the convergence of the energy components with increasing size of the naJC basis. A representative sample of such a convergence for $T_2$ is shown in Table 2. The convergence is guided by the shell parameter $\Omega$, which is defined as an upper bound to $\sum_{i=1}^{5} k_i$. As can be inferred from this table, the uncertainty assigned to the extrapolated energy is of the order of $10^{-14}$ a.u., which corresponds to $10^{-8}$ cm$^{-1}$ uncertainty in dissociation energy. This, however, is just the numerical uncertainty resulting from the finite size of the basis set used in the practical calculations. In fact, slightly larger uncertainty will be obtained if the limited accuracy of physical constants is taken into account. Uncertainties transferred to the energy from the Rydberg constant affect dis-sociation energy $D_{vJ}$ at the level of $10^{-8}$ cm$^{-1}$, while those from the nuclear masses can affect it even at $10^{-6}$ cm$^{-1}$. This comparison says that further diminishing of the numerical uncertainty, although potentially achievable by increasing the basis set size, will be reasonable only after the accuracy of the relevant physical constants is increased. Having in mind the future changes in values of the proton, $m_p$, deuteron, $m_d$, and triton, $m_t$, masses, we have evaluated the corresponding derivatives of the energy $\partial E_{vJ}/\partial m$. Their numerical values are available in Tables 3–5. Using these derivatives, the adjusted energy can then readily be obtained.

**Table 2.** Convergence of the nonrelativistic energy of the ground level $E_{00}$ (in a.u) and the corresponding dissociation energy $D_{00}$ (in cm$^{-1}$) with increasing basis set size $K$ for T$_2$.

| $\Omega$ | $K$ | $E_{00}$ | $D_{00}$ |
|---|---|---|---|
| 10 | 58 968 | −1.168 535 675 725 407 | 37 029.224 865 52 |
| 11 | 84 672 | −1.168 535 675 731 097 | 37 029.224 866 77 |
| 12 | 118 944 | −1.168 535 675 732 467 | 37 029.224 867 07 |
| 13 | 163 296 | −1.168 535 675 732 754 | 37 029.224 867 13 |
| 14 | 220 320 | −1.168 535 675 732 836 | 37 029.224 867 15 |
| $\infty$ | $\infty$ | −1.168 535 675 732 86(3) | 37 029.224 867 15(1) |

### 3.2. Nonrelativistic energy of HT, DT, and T$_2$

Tables 3–5 assemble the results for all three tritium-containing isotopologues of the hydrogen molecule. The first two columns identify the energy level with the vibrational $v$ and rotational $J$ quantum number. Third column contains the eigenvalue of the nonrelativistic Hamiltonian (1) extrapolated to the infinite basis size and accompanied by an estimated uncertainty. The next column shows the corresponding dissociation energy values obtained from

$$D_{vJ} = 2\mathcal{R}_\infty (E_\infty - E_{vJ}) \qquad (12)$$

where $\mathcal{R}_\infty$ is the Rydberg constant and $E_\infty$ the nonrelativistic energy of separated atoms. The latter energy was determined as a sum of the atomic ground state nonrelativistic energies (in a.u.)

$$E_\infty(AB) = E(A) + E(B) = -\frac{1}{2}\left(\frac{Z_A^2 m_A}{m_A + 1} + \frac{Z_B^2 m_B}{m_B + 1}\right). \qquad (13)$$

Apart from the DNA variational calculations reported in this work, the nonrelativistic energy can also be accurately evaluated using common methods based on the Born-Oppenheimer approximation. One practical implementation of such an approach is the nonadiabatic perturbation theory (NAPT) [49, 51, 56, 82, 83]. Within this perturbational method, the energy is evaluated as a series in powers of inverse reduced nuclear masses $1/\mu_n$. The cutoff of this series introduces a certain error in the energy. Thus far, the only implementation of the NAPT method involves the energy terms up to the second order [NAPT(2)]. The DNA method yields the energy, which can be viewed as an equivalent of the infinite perturbational series, and as such can be used to estimate the truncation error of the NAPT(2) method and to verify the uncertainties assigned to the nonrelativistic energy. So far, in NAPT(2) calculations the uncertainty of the nonrelativistic energy was estimated using a simple scaling of the leading-order nonadiabatic contribution by the $1/\mu_n$ factor. In view of the results quoted in the last column of Tables 3–5, this estimation turns out to be up to one order of magnitude too conservative.

The missing contribution to energy from higher-order ($n > 2$) terms of NAPT expansion behaves very regularly when going up the rotational ladder of states. Figure 1 in panels (a)-(c) shows the dependence of the missing contribution on the $J(J+1)$ factor. This relationship is almost perfectly linear, which enables reliable prediction of the error for higher $J$. Panel (d) in Fig. 1 illustrates the influence of the nuclear masses on the missing contribution; for the lightest isotopologue, HT, this contribution

**Table 3.** Nonrelativistic energy and the corresponding dissociation energy of selected rovibrational levels $(v, J)$ of HT. The uncertainties assigned to $E_{vJ}$ and $D_{vJ}$ are due to the numerical convergence only and do not account for uncertainties transferred from the fundamental constants. $\partial E_{vJ}/\partial m_t$ and $\partial E_{vJ}/\partial m_p$ (in a.u.) are derivatives of the energy with respect to the mass of the triton and proton, respectively. The last column shows the contribution to $D_{vJ}$ from the higher order terms of the NAPT (see text for discussion).

| $v$ | $J$ | $E_{vJ}$/a.u. | $D_{vJ}$/cm$^{-1}$ | $\dfrac{\partial E_{vJ}}{\partial m_t} \times 10^7$ | $\dfrac{\partial E_{vJ}}{\partial m_p} \times 10^7$ | $\delta D_{vJ}^{\mathrm{NAPT}(n>2)}$/cm$^{-1}$ |
|---|---|---|---|---|---|---|
| 0 | 0 | $-1.166\,002\,037\,328\,64(3)$ | $36\,512.928\,009\,11(1)$ | $-2.00$ | $-17.9$ | $0.000\,006\,96$ |
| 0 | 1 | $-1.165\,640\,028\,983\,64(4)$ | $36\,433.476\,361\,04(1)$ | $-2.15$ | $-19.3$ | $0.000\,012\,80$ |
| 0 | 2 | $-1.164\,918\,229\,777\,66(3)$ | $36\,275.059\,746\,39(1)$ | $-2.48$ | $-22.2$ | $0.000\,024\,45$ |
| 0 | 3 | $-1.163\,841\,027\,541\,52(3)$ | $36\,038.641\,182\,71(1)$ | $-2.96$ | $-26.5$ | $0.000\,041\,78$ |
| 0 | 4 | $-1.162\,414\,888\,556\,04(4)$ | $35\,725.639\,854\,60(1)$ | $-3.59$ | $-32.2$ | $0.000\,064\,65$ |
| 0 | 5 | $-1.160\,648\,225\,433\,93(5)$ | $35\,337.902\,117\,13(1)$ | $-4.37$ | $-39.1$ | $0.000\,092\,82$ |
| 1 | 0 | $-1.150\,351\,886\,701\,26(4)$ | $33\,078.116\,969\,39(1)$ | $-5.39$ | $-48.3$ | $0.000\,094\,07$ |
| 1 | 1 | $-1.150\,004\,630\,468\,46(5)$ | $33\,001.903\,035\,70(1)$ | $-5.54$ | $-49.7$ | $0.000\,098\,58$ |
| 1 | 2 | $-1.149\,312\,274\,792\,04(5)$ | $32\,849.948\,528\,85(1)$ | $-5.85$ | $-52.4$ | $0.000\,107\,57$ |
| 1 | 3 | $-1.148\,279\,086\,759\,89(5)$ | $32\,623.189\,966\,36(1)$ | $-6.29$ | $-56.4$ | $0.000\,120\,95$ |
| 1 | 4 | $-1.146\,911\,353\,189\,40(6)$ | $32\,323.007\,145\,18(2)$ | $-6.89$ | $-61.7$ | $0.000\,138\,62$ |
| 1 | 5 | $-1.145\,217\,250\,575\,04(6)$ | $31\,951.194\,598\,40(1)$ | $-7.62$ | $-68.2$ | $0.000\,160\,40$ |
| 2 | 0 | $-1.135\,422\,043\,069\,87(6)$ | $29\,801.395\,042\,08(1)$ | $-8.46$ | $-75.8$ | $0.000\,146\,31$ |
| 2 | 1 | $-1.135\,089\,253\,849\,01(6)$ | $29\,728.356\,250\,51(1)$ | $-8.61$ | $-77.1$ | $0.000\,149\,90$ |
| 2 | 2 | $-1.134\,425\,774\,109\,68(7)$ | $29\,582.739\,279\,30(1)$ | $-8.89$ | $-79.7$ | $0.000\,157\,06$ |
| 2 | 3 | $-1.133\,435\,755\,481\,11(7)$ | $29\,365.455\,305\,75(1)$ | $-9.31$ | $-83.4$ | $0.000\,167\,72$ |
| 2 | 4 | $-1.132\,125\,313\,406\,32(7)$ | $29\,077.846\,514\,46(2)$ | $-9.87$ | $-88.4$ | $0.000\,181\,80$ |
| 2 | 5 | $-1.130\,502\,399\,328\,49(8)$ | $28\,721.658\,045\,50(2)$ | $-10.5$ | $-94.5$ | $0.000\,199\,17$ |

**Table 4.** Nonrelativistic energy and the corresponding dissociation energy of selected rovibrational levels $(v, J)$ of DT. The uncertainties assigned to $E_{vJ}$ and $D_{vJ}$ are due to the numerical convergence only and do not account for uncertainties transferred from the fundamental constants. $\partial E_{vJ}/\partial m_t$ and $\partial E_{vJ}/\partial m_d$ (in a.u.) are derivatives of the energy with respect to the mass of the triton and deuteron, respectively. The last column shows the contribution to $D_{vJ}$ from the higher order terms of the NAPT (see text for discussion).

| $v$ | $J$ | $E_{vJ}$/a.u. | $D_{vJ}$/cm$^{-1}$ | $\dfrac{\partial E_{vJ}}{\partial m_t} \times 10^7$ | $\dfrac{\partial E_{vJ}}{\partial m_d} \times 10^7$ | $\delta D_{vJ}^{\mathrm{NAPT}(n>2)}$/cm$^{-1}$ |
|---|---|---|---|---|---|---|
| 0 | 0 | $-1.167\,819\,673\,436\,73(3)$ | $36\,882.009\,843\,48(1)$ | $-2.48$ | $-5.59$ | $0.000\,006\,56$ |
| 0 | 1 | $-1.167\,592\,154\,385\,95(3)$ | $36\,832.075\,183\,68(1)$ | $-2.66$ | $-5.94$ | $0.000\,006\,89$ |
| 0 | 2 | $-1.167\,137\,988\,265\,85(3)$ | $36\,732.397\,241\,90(1)$ | $-2.99$ | $-6.70$ | $0.000\,007\,56$ |
| 0 | 3 | $-1.166\,458\,907\,411\,88(3)$ | $36\,583.356\,221\,81(1)$ | $-3.46$ | $-7.77$ | $0.000\,008\,55$ |
| 0 | 4 | $-1.165\,557\,481\,627\,24(3)$ | $36\,385.516\,130\,02(1)$ | $-4.11$ | $-9.22$ | $0.000\,009\,88$ |
| 0 | 5 | $-1.164\,437\,084\,754\,32(3)$ | $36\,139.617\,439\,36(1)$ | $-4.91$ | $-11.0$ | $0.000\,011\,55$ |
| 1 | 0 | $-1.155\,320\,098\,608\,18(4)$ | $34\,138.670\,265\,79(1)$ | $-6.87$ | $-15.4$ | $0.000\,022\,55$ |
| 1 | 1 | $-1.155\,099\,913\,443\,84(4)$ | $34\,090.345\,208\,01(1)$ | $-7.02$ | $-15.8$ | $0.000\,022\,79$ |
| 1 | 2 | $-1.154\,660\,396\,129\,22(4)$ | $33\,993.882\,307\,41(1)$ | $-7.34$ | $-16.5$ | $0.000\,023\,28$ |
| 1 | 3 | $-1.154\,003\,241\,186\,62(4)$ | $33\,849.653\,468\,63(1)$ | $-7.80$ | $-17.5$ | $0.000\,024\,02$ |
| 1 | 4 | $-1.153\,130\,962\,007\,59(4)$ | $33\,658.210\,317\,37(1)$ | $-8.41$ | $-18.9$ | $0.000\,025\,02$ |
| 1 | 5 | $-1.152\,046\,857\,806\,40(4)$ | $33\,420.276\,947\,45(1)$ | $-9.17$ | $-20.6$ | $0.000\,026\,26$ |
| 2 | 0 | $-1.143\,274\,252\,183\,51(5)$ | $31\,494.912\,562\,28(1)$ | $-10.9$ | $-24.5$ | $0.000\,034\,01$ |
| 2 | 1 | $-1.143\,061\,281\,369\,56(5)$ | $31\,448.170\,871\,39(1)$ | $-11.1$ | $-24.8$ | $0.000\,034\,17$ |
| 2 | 2 | $-1.142\,636\,174\,416\,21(5)$ | $31\,354.870\,679\,52(1)$ | $-11.4$ | $-25.5$ | $0.000\,034\,53$ |
| 2 | 3 | $-1.142\,000\,589\,298\,66(5)$ | $31\,215.375\,870\,14(1)$ | $-11.8$ | $-26.5$ | $0.000\,035\,04$ |
| 2 | 4 | $-1.141\,156\,984\,922\,24(5)$ | $31\,030.226\,110\,61(1)$ | $-12.4$ | $-27.8$ | $0.000\,035\,75$ |
| 2 | 5 | $-1.140\,108\,588\,485\,95(5)$ | $30\,800.129\,689\,24(1)$ | $-13.1$ | $-29.4$ | $0.000\,036\,63$ |

**Table 5.** Nonrelativistic energy and the corresponding dissociation energy of selected rovibrational levels $(v, J)$ of $T_2$. The uncertainties assigned to $E_{vJ}$ and $D_{vJ}$ are due to the numerical convergence only and do not account for uncertainties transferred from the fundamental constants. $\partial E_{vJ}/\partial m_t$ (in a.u.) is the derivative of the energy with respect to the mass of the triton. The last column shows the contribution to $D_{vJ}$ from the higher order terms of the NAPT (see text for discussion).

| $v$ | $J$ | $E_{vJ}$/a.u. | $D_{vJ}$/cm$^{-1}$ | $\dfrac{\partial E_{vJ}}{\partial m_t} \times 10^7$ | $\delta D_{vJ}^{\mathrm{NAPT}(n>2)}$/cm$^{-1}$ |
|---|---|---|---|---|---|
| 0 | 0 | −1.168 535 675 732 86(3) | 37 029.224 867 15(1) | −5.53 | 0.000 004 49 |
| 0 | 1 | −1.168 353 137 671 73(3) | 36 989.162 393 48(1) | −5.86 | 0.000 004 62 |
| 0 | 2 | −1.167 988 621 785 72(3) | 36 909.160 403 77(1) | −6.51 | 0.000 004 88 |
| 0 | 3 | −1.167 443 242 559 83(3) | 36 789.463 499 21(1) | −7.49 | 0.000 005 26 |
| 0 | 4 | −1.166 718 656 905 26(3) | 36 630.435 329 79(1) | −8.79 | 0.000 005 78 |
| 0 | 5 | −1.165 817 046 799 18(3) | 36 432.554 784 12(1) | −10.4 | 0.000 006 43 |
| 1 | 0 | −1.157 306 578 041 28(4) | 34 564.722 790 75(1) | −15.4 | 0.000 014 40 |
| 1 | 1 | −1.157 129 306 019 02(4) | 34 525.816 079 02(1) | −15.7 | 0.000 014 50 |
| 1 | 2 | −1.156 775 311 284 55(4) | 34 448.123 215 16(1) | −16.3 | 0.000 014 71 |
| 1 | 3 | −1.156 245 686 525 60(4) | 34 331.884 016 43(1) | −17.3 | 0.000 015 02 |
| 1 | 4 | −1.155 542 056 097 14(3) | 34 177.454 987 53(1) | −18.5 | 0.000 015 44 |
| 1 | 5 | −1.154 666 558 833 37(5) | 33 985.305 548 30(1) | −20.1 | 0.000 015 96 |
| 2 | 0 | −1.146 441 884 581 32(5) | 32 180.198 198 75(1) | −24.6 | 0.000 022 41 |
| 2 | 1 | −1.146 269 799 929 89(4) | 32 142.429 983 32(1) | −24.9 | 0.000 022 49 |
| 2 | 2 | −1.145 926 169 331 39(4) | 32 067.011 784 39(1) | −25.5 | 0.000 022 65 |
| 2 | 3 | −1.145 412 064 321 78(4) | 31 954.178 776 92(1) | −26.4 | 0.000 022 88 |
| 2 | 4 | −1.144 729 077 678 78(5) | 31 804.280 535 22(1) | −27.6 | 0.000 023 20 |
| 2 | 5 | −1.143 879 306 418 14(5) | 31 617.777 301 05(1) | −29.1 | 0.000 023 59 |

grows much faster with $J$ than for the other, heavier molecules. This observation is consistent with the intuition saying that the heavier the nuclei, the more accurate the results obtained in the framework of the clamped nuclei approximation.

### 3.3. Comparison to literature data

Current DNA calculations using the naJC wavefunction exceed by several orders of magnitude the best previous calculations reported in the literature. To illustrate this claim, we compare in Table 6 the nonrelativistic energy of the lowest rotational and vibrational levels of $T_2$ obtained with different methods. This comparison includes the direct variational calculations with explicitly correlated Gaussian (ECG) wavefunctions by Adamowicz et al. [50, 53], as well as calculations performed in the framework of the Born-Oppenheimer approximation by Wolniewicz [47] and by us [51]. Among them, the most accurate are the NAPT calculations [51] improving by ca. five orders of magnitude the results obtained by Wolniewicz in 1984 [47]. For the ground rovibrational level $(v = 0, J = 0)$ also the ECG calculations [50] give accurate energy. However, for $J = 1$ the accuracy of the ECG-based calculations [53] decreases by two orders of magnitude. The error of the other two methods does not depend on $J$.

### 3.4. Comparison to experimental data

Quite recently, measurements of fundamental vibrational splittings of selected transitions in the tritiated isotopologues of $H_2$ have been reported [44]. For Q($J$) lines, the absolute accuracy of $5 \cdot 10^{-4}$ cm$^{-1}$ has been attained, in agreement with slightly more accurate theoretical data from NAPT calculations [55]. The uncertainty of these theoretical results was dominated by the contribution from the nonrelativistic energy. Our new DNA calculations eliminated this contribution and diminished the overall uncertainty. The experimental frequencies are juxtaposed against theoretical data in

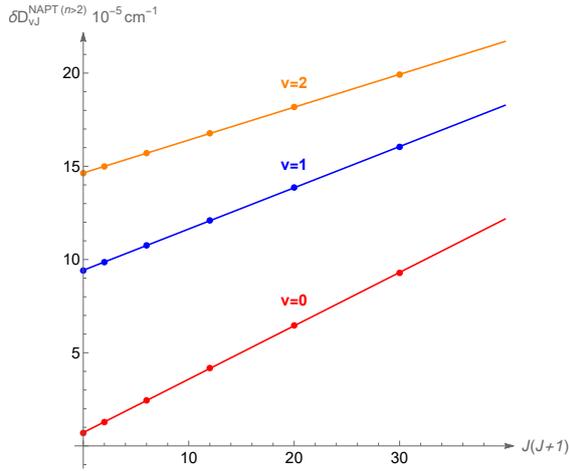
(a) HT, $v = 0, 1, 2$.

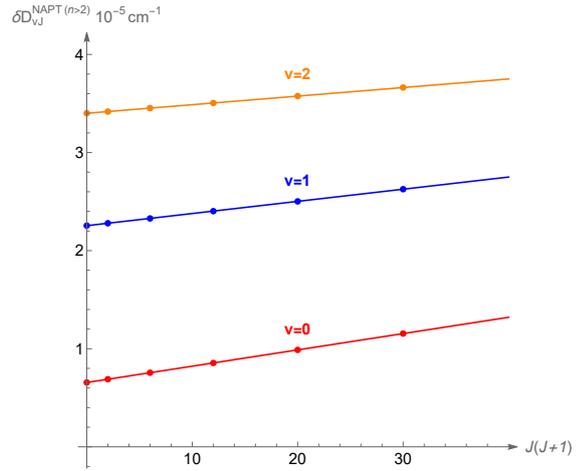
(b) DT, $v = 0, 1, 2$.

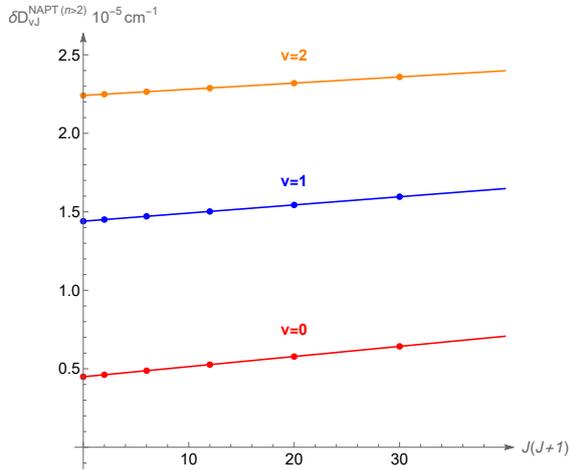
(c) T$_2$, $v = 0, 1, 2$.

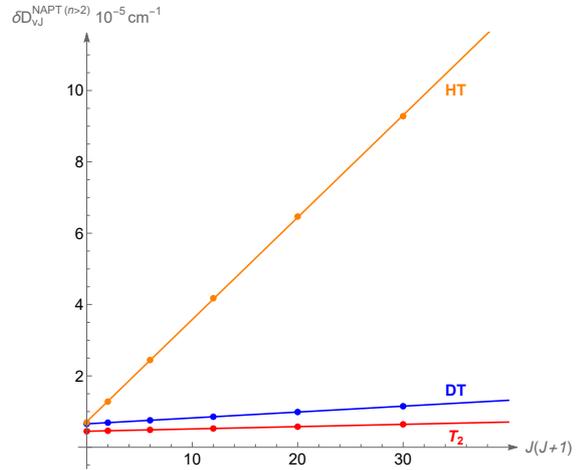
(d) $v = 0$, HT, DT, T$_2$.

**Figure 1.** Contribution from higher-order ($n > 2$) terms of NAPT expansion to the dissociation energy of the rovibrational levels $J = 0 - 5$ in the $v = 0, 1, 2$ states for (a) HD, (b) DT, and (c) T$_2$. Panel (d) shows the contribution for different rotational levels at the ground vibrational state, $v = 0$, for all three molecules.

Table 7. By comparing the uncertainties in columns 'NAPT' and 'DNA+NAPT', one can conclude that the total uncertainty of the transition energies is now 28 times smaller for HT, 11 times smaller for DT, and 7 times smaller for T$_2$. At present, the leading contribution to the error budget comes from the quantum electrodynamic correction and the total theoretical uncertainty is on the order of $10^{-5}$ cm$^{-1}$.

**Table 6.** Comparison of the nonrelativistic energy $E_{v,J}$ of T$_2$ (in atomic units) obtained from the DNA method to the best literature data $E_{vJ}^{\text{lit}}$ of the selected lowest vibrational levels reported by Bubin *et al.* [50], Kirnosov *et al.* [53], Pachucki *et al.* [51], and by Wolniewicz *et al.* [47]. $\delta E_{vJ} = E_{vJ} - E_{vJ}^{\text{lit}}$.

| $v$ | $E_{vJ}$ | $E_{vJ}^{\text{lit}}$ [50, 53] | $10^{11}\,\delta E_{vJ}$ | $E_{vJ}^{\text{lit}}$ [51] | $10^{11}\,\delta E_{vJ}$ | $E_{vJ}^{\text{lit}}$ [47] | $10^{8}\,\delta E_{vJ}$ |
|---|---|---|---|---|---|---|---|
| | | | $J=0$ | | | | |
| 0 | −1.168 535 675 732 86(3) | −1.168 535 675 71 | −2 | −1.168 535 675 71 | −2 | −1.168 531 80 | −387 |
| 1 | −1.157 306 578 041 28(4) | −1.157 306 577 87 | −17 | −1.157 306 577 98 | −7 | −1.157 301 80 | −478 |
| 2 | −1.146 441 884 581 32(5) | −1.146 441 883 89 | −69 | −1.146 441 884 48 | −10 | −1.146 436 31 | −558 |
| | | | $J=1$ | | | | |
| 0 | −1.168 353 137 671 73(3) | −1.168 353 133 98 | −369 | −1.168 353 137 65 | −2 | −1.168 349 26 | −388 |
| 1 | −1.157 129 306 019 02(4) | −1.157 129 302 16 | −386 | −1.157 129 305 95 | −7 | −1.157 124 52 | −478 |
| 2 | −1.146 269 799 929 89(4) | −1.146 269 795 55 | −438 | −1.146 269 799 83 | −10 | −1.146 264 22 | −558 |
| | | | $J=2$ | | | | |
| 0 | −1.167 988 621 785 72(3) | − | | −1.167 988 621 76 | −2 | −1.167 984 73 | −389 |
| 1 | −1.156 775 311 284 55(4) | − | | −1.156 775 311 22 | −7 | −1.156 770 51 | −480 |
| 2 | −1.145 926 169 331 39(4) | − | | −1.145 926 169 23 | −10 | −1.145 920 58 | −559 |
| | | | $J=3$ | | | | |
| 0 | −1.167 443 242 559 83(3) | − | | −1.167 443 242 54 | −2 | −1.167 439 32 | −392 |
| 1 | −1.156 245 686 525 60(4) | − | | −1.156 245 686 46 | −7 | −1.156 240 91 | −478 |
| 2 | −1.145 412 064 321 78(4) | − | | −1.145 412 064 22 | −10 | −1.145 406 45 | −562 |

**Table 7.** Comparison of the most accurate experimental data [44] for the fundamental $v = 0 \to 1$ vibrational splittings of selected transitions in HT, DT, and T$_2$ with theoretical data obtained from NAPT calculations [55] and from the new DNA calculations of the nonrelativistic energy augmented by relativistic and quantum-electrodynamic corrections from NAPT [55, 56] (in cm$^{-1}$). The 'Difference' concerns the latter calculations and the measurements.

| Line | Experiment [44] | NAPT [55] | DNA+NAPT | Difference |
|---|---|---|---|---|
| | | HT | | |
| Q(0) | 3 434.812 48(53) | 3 434.813 33(44) | 3 434.813 253(16) | −0.000 77(53) |
| Q(1) | 3 431.575 09(53) | 3 431.575 53(44) | 3 431.575 453(16) | −0.000 36(53) |
| Q(2) | 3 425.112 65(53) | 3 425.113 24(44) | 3 425.113 173(16) | −0.000 52(53) |
| Q(3) | 3 415.452 58(53) | 3 415.452 98(44) | 3 415.452 914(16) | −0.000 33(53) |
| | | DT | | |
| Q(0) | 2 743.341 60(42) | 2 743.341 74(11) | 2 743.341 724(10) | −0.000 12(42) |
| Q(1) | 2 741.732 04(39) | 2 741.732 10(11) | 2 741.732 079(10) | −0.000 04(39) |
| Q(2) | 2 738.516 62(42) | 2 738.516 97(11) | 2 738.516 953(10) | −0.000 33(42) |
| Q(3) | 2 733.704 79(42) | 2 733.704 66(11) | 2 733.704 645(10) | +0.000 15(42) |
| Q(4) | 2 727.307 45(42) | 2 727.307 55(11) | 2 727.307 535(10) | −0.000 09(42) |
| Q(5) | 2 719.342 21(42) | 2 719.342 02(11) | 2 719.342 005(10) | +0.000 20(42) |
| | | T$_2$ | | |
| Q(0) | 2 464.503 94(67) | 2 464.504 15(6) | 2 464.504 142(8) | −0.000 20(67) |
| Q(1) | 2 463.348 17(42) | 2 463.348 36(6) | 2 463.348 350(8) | −0.000 18(42) |
| Q(2) | 2 461.039 17(42) | 2 461.039 17(6) | 2 461.039 163(8) | +0.000 01(42) |
| Q(3) | 2 457.581 35(42) | 2 457.581 37(6) | 2 457.581 366(8) | −0.000 02(42) |
| Q(4) | 2 452.982 33(42) | 2 452.982 11(6) | 2 452.982 104(8) | +0.000 23(42) |
| Q(5) | 2 447.250 61(42) | 2 447.250 85(6) | 2 447.250 847(8) | −0.000 24(42) |
| S(0) | 2 581.114(5) | 2 581.105 22(6) | 2 581.105 213(9) | +0.009(5) |
| S(1) | 2 657.281(5) | 2 657.282 90(6) | 2 657.282 890(9) | −0.002(5) |
| S(2) | 2 731.716(5) | 2 731.710 84(6) | 2 731.710 828(10) | +0.005(5) |
| S(3) | 2 804.164(5) | 2 804.164 21(6) | 2 804.164 200(10) | −0.000(5) |

## 4. Conclusion

The direct nonadiabatic (DNA) variational method in connection with the nonadiabatic James-Coolidge (naJC) wavefunction enables the nonrelativistic energy of hydrogen molecule isotopologues to be determined with a relative accuracy of $10^{-13}-10^{-14}$. This method preserves its accuracy also for rotationally and vibrationally excited states. In terms of the dissociation energy, its nonrelativistic component is accurate to $10^{-7} - 10^{-8}\,\text{cm}^{-1}$, which surpasses the uncertainty due to inaccuracy in the nuclear masses. Now, the contribution of the nonrelativistic energy to the overall uncertainty budget can be safely neglected because it is several orders of magnitude smaller than the remaining contributions originating from relativistic and quantum-electrodynamic corrections. In effect, the final accuracy of transition energies increased by an order of magnitude, and it is justified now to focus on improving the accuracy of subsequent energy components.

Variational energies obtained in this work will be included in the public H2Spectre program [55] as internal data.

**Disclosure statement**

We declare no conflict of interest.

**Funding**

This research was supported by National Science Center (Poland) Grant No. 2017/25/B/ST4/01024, as well as by a computing grant from the Poznan Supercomputing and Networking Center.